\documentclass[twoside]{dis09}
\usepackage[latin1]{inputenc}
\usepackage[dvips]{graphicx,epsfig,color}
\usepackage{wrapfig,rotating}
\usepackage{amssymb,amsmath,array}
\usepackage{subfigure}

\begin{document}
\title{Cross-sections and Single Spin Asymmetries \\
of Identified Hadrons in $p$$^\uparrow$+$p$ at $\sqrt{s}$ = 200 GeV}

\author{J.H. Lee and F. Videb{\ae}k for the BRAHMS Collaboration
%
\thanks{This work was supported by Brookhaven Science Associates, LLC under Contract
No. DE-AC02-98CH10886 with the U.S. Department of Energy.}
%
\vspace{.3cm}\\
%
Brookhaven National Laboratory - Physics Department \\
Upton, NY 11973 - U.S.A.
}

\maketitle

\begin{abstract}
Measurements of cross-sections and transverse single spin asymmetries of
identified charged hadrons at forward rapidities from transversely
polarized proton collisions at $\sqrt{s}$ = 200 GeV are presented.  
The results are discussed in the context of interplay between 
perturbative and non-perturbative QCD effects.
\end{abstract}

\section{Introduction}
Despite decades of experimental and theoretical efforts, the exact nature
of spin structure of the nucleon still remains elusive. One of the puzzles is 
the origin of the large transverse Single Spin Asymmetries (SSAs) at large-$x_F$(=$2p_L/\sqrt{s}$)
observed in a wide range of energies in $p^\uparrow +p$ ($\bar{p}^\uparrow+p$)
collisions~\cite{rev}, whereas SSAs are expected to be suppressed and 
may rise only as an ${\mathcal O}(1/p_T)$ effect at leading-twist 
in massless perturbative QCD (pQCD)~\cite{kane}.
Recently, new measurements of SSAs have been available 
from semi-inclusive deep-inelastic scattering (SIDIS)~\cite{hermes,compass} 
and  $p^\uparrow$+$p$ at RHIC~\cite{star,prl}.  They provide more opportunities for stringent tests for the 
pQCD inspired theoretical models to get insight into the fundamental mechanisms of SSA as well 
as the relevant hadron structure.
Main theoretical focuses to account for
the recently observed SSAs at high energies in the context of pQCD  have been on the role of 
partonic transverse momentum effects in the structure of the initial transversely
polarized nucleon~\cite{sivers}, and the fragmentation process of a polarized quark into
hadrons~\cite{collins}.  Higher twist effects arising from  quark-gluon correlations have
also been considered as a possible origin of SSA~\cite{qiu}. 
  These theoretical models have been  successful in
describing some aspects of SSAs, but there still remains challenges of characterizing all
transverse spin phenomena in the context of pQCD, which is in good agreements with unpolarized or spin-averaged 
cross-sections at RHIC. In spite of the theoretical progresses and efforts, it is still conceivable
that the spin degree of freedom even at the RHIC energy 
domain, where $\sqrt{s}$ $>\!\!>$ $\Lambda_{QCD}$, is not dominated by pQCD phenomena 
and substantially driven by contribution from non-pQCD effects~\cite{nonpqcd}. 

In these proceedings, we discuss the SSA measurements by BRAHMS at forward rapidities 
covering high-$x_F$ where large SSAs have been observed in 
$p$$^\uparrow$+$p$ at $\sqrt{s} =200$ GeV at RHIC. The discussion is focused on
$p_T$, flavor,  multiplicity, and energy dependence of SSAs in the context of possible 
signatures of non-pQCD effect of transverse spin phenomenon. 

\section{Validation of pQCD at RHIC: Spin-averaged cross-sections}
\begin{wrapfigure}{r}{0.68\columnwidth}
\centerline{\includegraphics[width=0.6\columnwidth]{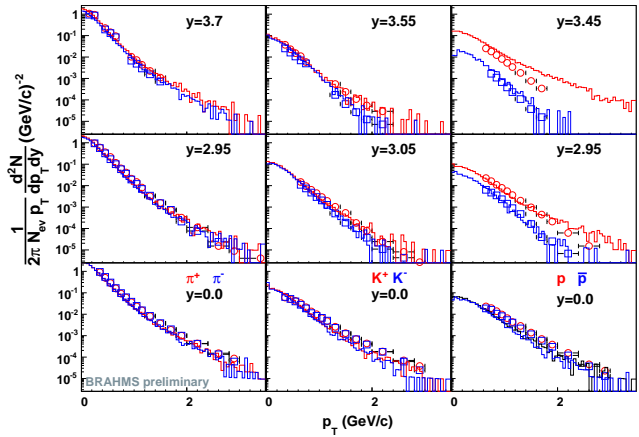}}
\caption{Invariant spectra for $\pi^{\pm}$ (left), $K^{\pm}$ (middle), and $p,\bar{p}$ (right)  
at 3 different rapidities indicated in each panel at $\sqrt{s}=200$ GeV. Positive and negative 
particles are displayed with circles and squares, respectively.
The spectra are compared with PYTHIA calculations.}\label{Fig:pythia}
\end{wrapfigure}

The invariant identified charged hadron spectra in $p+p$ at $\sqrt{s}$=200 GeV are shown in
Fig.~\ref{Fig:pythia}. 
The spectra are normalized to $\approx \rm{41 mb}$ of the total inelastic cross-section.
The data have been corrected for the geometrical acceptance of the spectrometer,
multiple scattering, loss of tracks due to weak decays,  and absorption in the material along
the path of the detected particles.
The spectra are not corrected for feed-down from
the weak decays of $K^0_s$, $\Lambda, \bar{\Lambda}$ and higher mass hyperons mainly due to
unknown hyperon yields at forward rapidities.
In Fig.~\ref{Fig:pythia}, data are also compared with the QCD-inspired Monte-Carlo model, 
PYTHIA~\cite{pythia}. 
The data and PYTHIA descriptions for the mesons in the all measured rapidities are in a good agreement.
A significant disagreement in describing 
protons at forward rapidities calls for better theoretical understanding of the baryon 
production mechanism, in particular the transport mechanism since a significant fraction 
of the protons in the fragmentation region might still be related to the protons transported
at the measured kinematic range.
Comparisons to the other pQCD calculations, including next-to-leading-order pQCD calculations to the BRAHMS
measurements have been made previously~\cite{brahms_pp,dss}. The comparisons have shown that the calculations 
for the meson production are also in good agreements down to $p_T$$\sim$1 GeV/$c$ to the data  
at mid- and forward rapidities. That confirms that in this energy regime pQCD is applicable for the un-polarized particle 
productions at RHIC in particular at high-rapidity region where large SSAs have been measured. 

\section{Single Spin Asymmetry Measurements at high-$x_F$}
The SSA in $p+p$ collisions is defined as a ``left-right'' asymmetry of produced particles from
the hadronic scattering of transversely polarized protons, and is customary defined as 
analyzing power $A_N$: 
\begin{equation} 
A_N \equiv \frac{1}{\mathcal P} \frac{(N^+ - {\mathcal L}N^-)}{(N^+ + {\mathcal L}N^-)}, 
\label{eq:An}
\end{equation}
where ${\mathcal P}$ is the polarization of the beam, ${\mathcal L}$ is the
spin dependent relative luminosity (${\mathcal L}$ = ${\mathcal
  L_+}$/${\mathcal L_-}$)
and $N^{+(-)}$ is the number of detected particles with beam spin
vector oriented up (down).  
The average polarization of the beam ${\mathcal P}$ as
determined from the CNI measurements is about $50\%$ for RHIC Run-5.
The systematic error on the $A_N$ measurements is estimated to be $\sim$$10\%$ including
uncertainties from the beam polarization ($\sim$6$\%$).  
The systematic error represents mainly scaling uncertainties on the values of $A_N$.
The data presented here were collected with the BRAHMS detector
system~\cite{brahms_nim} in polarized $p+p$ collisions from Run-5 with recorded integrated luminosity
corresponding to 2.4 pb$^{-1}$ at $\sqrt{s}=200$ GeV.
The kinematic coverage of the data taken with BRAHMS Forward Spectrometer at 2.3$^\circ$ and 4$^\circ$
for  $\sqrt{s}=200$ GeV as a function of $p_T$ and $x_F$ can be found in~\cite{dis07}. 

\begin{wrapfigure}{r}{0.7\columnwidth}
\centerline{\includegraphics[width=0.65\columnwidth]{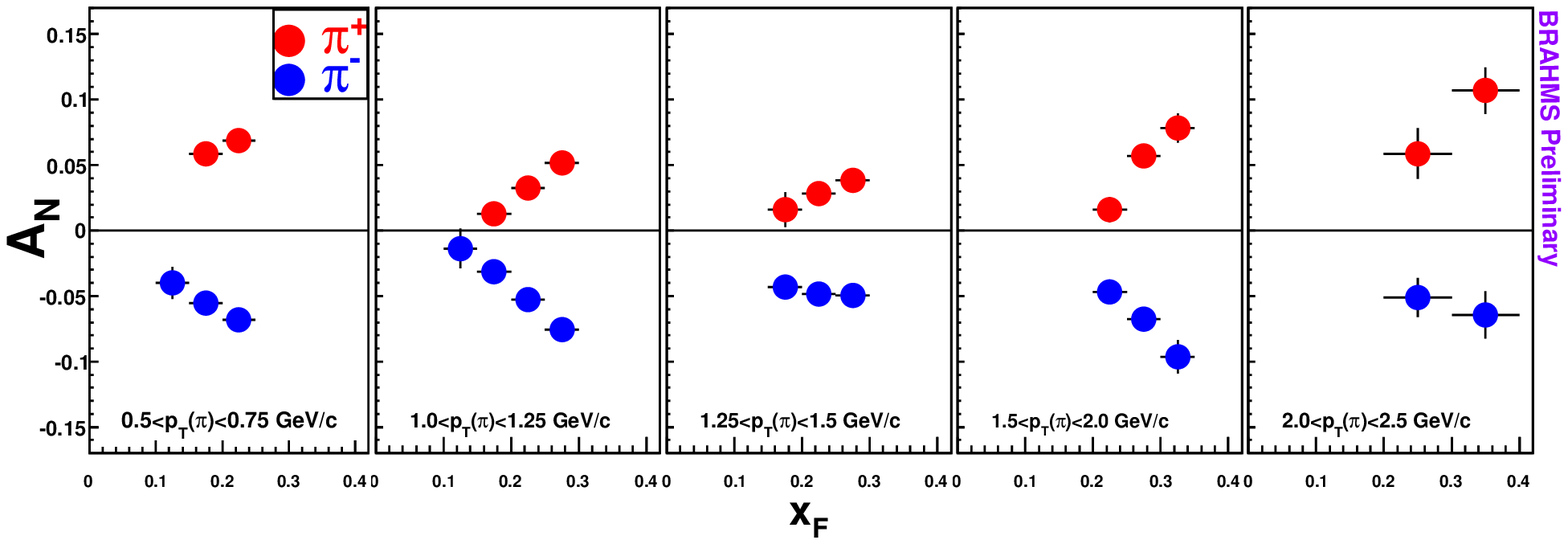}}
\caption{$A_N$  vs. $x_F$  for $\pi^+$ and $\pi^-$ at $\sqrt{s}$ = 200 GeV
at fixed $p_T$ values:
$0.5<p_T<0.75$, $1.<p_T<1.25$, $1.25<p_T<1.5$, $1.5<p_T<2.0$,  and $2.0<p_T<2.5$ GeV/$c$, respectively.}\label{Fig:pt_dep}
\end{wrapfigure}

\section{$p_T$-dependent Single Spin Asymmetries}
One of the strong indications that SSAs are in accordance with pQCD description can be  
power-suppressed nature of $A_N$, which
should be realized in the data as an decrease of SSAs with $p_T$.
Fig.~\ref{Fig:pt_dep} shows $A_N(\pi^+)$ and $A_N(\pi^-)$ as a function of $x_F$ for 5 different
$p_T$ regions from 0.5 to 2.5 GeV/$c$.   
The measurements exhibit no clear systematic $p_T$-dependence of SSAs in the measured kinematic range. 
The absence of consistent $1/p_T$-dependence in the data is similar to what was reported by STAR 
in $p+p\rightarrow \pi^0+X$~\cite{star_ptdep}
in the same reaction. It is possibly due to a non-trivial interplay between soft and hard processes 
in the measured kinematic region complicated by limited acceptance coverage of the experiment. 

\begin{wrapfigure}{r}{0.5\columnwidth}
\centerline{\includegraphics[width=0.47\columnwidth]{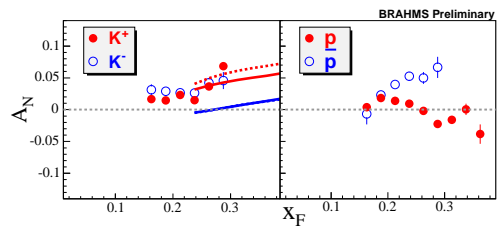}}
\caption{$A_N$ vs. $x_F$ for $K^{\pm}$ (left), and $p$ and $\bar{p}$ (right) at $\sqrt{s}$ $=$ 200 GeV. 
The curves are from the 
twist-3 calculations with (solid line) and without (broken line) sea- and anti-quark contribution.
}\label{Fig:k_p}
\end{wrapfigure}
\section{Flavor-dependent Single Spin Asymmetries}
Since partonic description of SSAs are expected to be sensitive mainly to the 
valence content of the particles measured, flavor dependence of asymmetries will provide strong constraints
to the theoretical description of SSAs. 
The SSAs for $K^{\pm}$, and $p,\bar{p}$
as a function of $x_F$ are shown in Fig.~\ref{Fig:k_p}.
The asymmetries for $K^+(u\bar{s})$ is positive as is the $A_N$ of $\pi^+(u\bar{d})$, which is expected
if the asymmetry is mainly carried by valence quarks, but the measured positive SSAs of $K^-(\bar{u}s)$ seem to
contradict  the n\"{a}ive expectations~\cite{ansel} of valence quark dominance in the SSA producing mechanism.
In a valence-like model (no Sivers effect from sea-quarks and/or gluons),
non-zero positive  $A_N(K^-)$ implies large non-leading fragmentation
functions ($D_u^{K^-}$, $D_d^{K^-}$) and insignificant contribution from strange quarks.
Twist-3 calculations~\cite{feng} shown in the figure also under-predict $A_N(K^-)$ 
due to the small contribution of sea and strange-quark contribution to $A_N$ in the model.  
This non-vanishing SSAs for $K^-$ is consistent with the measurements  in $p+p$ at $\sqrt{s} = 62.4$ GeV~\cite{prl},
but not with the preliminary results by HERMES, where large SSAs for $K^+$ were observed while SSAs for $K^-$ 
were consistent with zero in SIDIS~\cite{hermes_k}.
In Fig.~\ref{Fig:k_p}, anti-protons which carry no valence quark show also significant positive $A_N$ as for $K^-$, but 
protons show no significant asymmetries
in contrast to pions and kaons in the same kinematic region.  The insignificant observed asymmetries of protons
are consistent with the measurements at lower energies~\cite{saroff,40gev_proton}.
\begin{figure}[htp]
  \begin{center}
    \subfigure[]{\label{Fig:rpp}\includegraphics[scale=0.28]{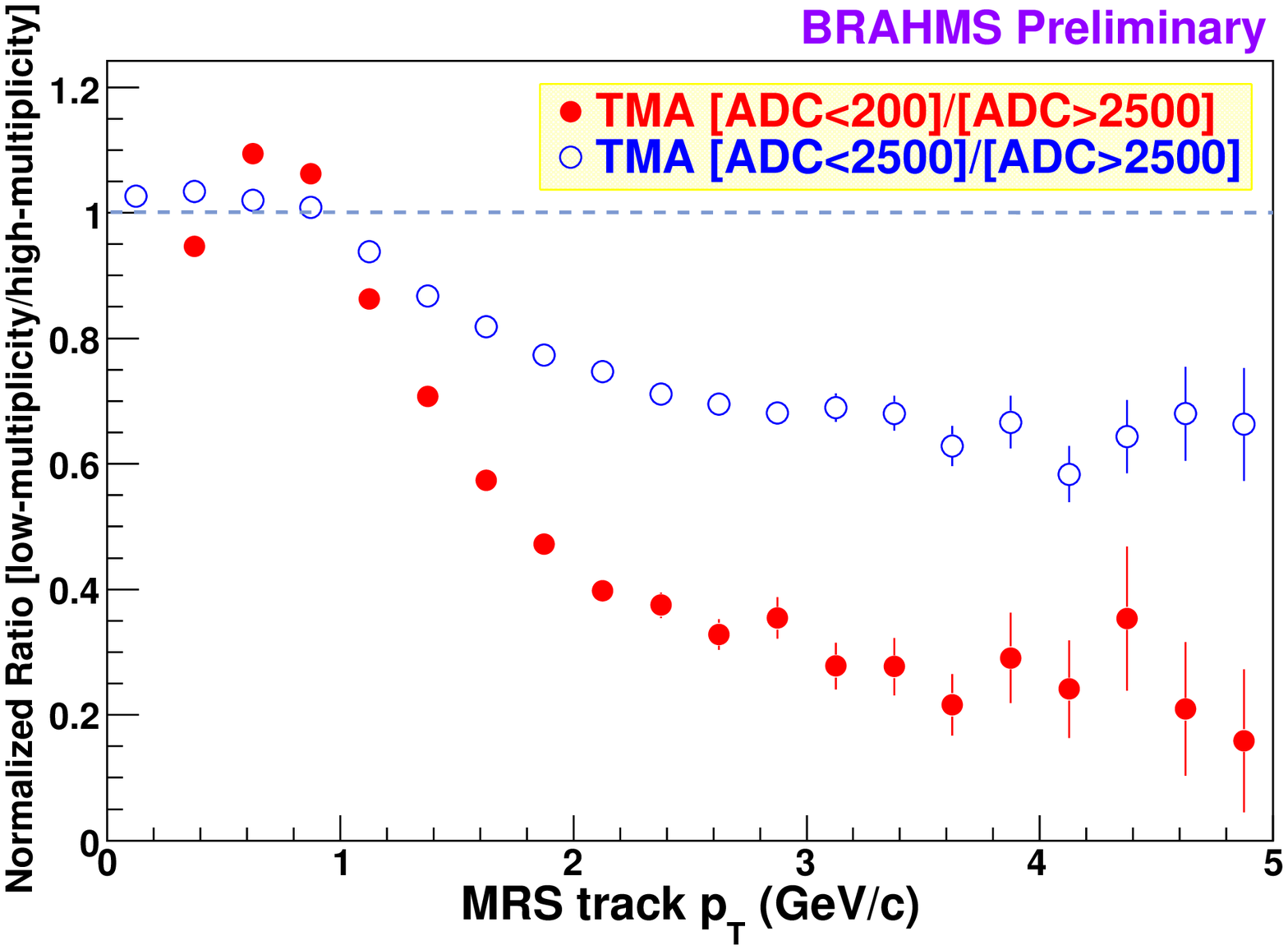}}
    \subfigure[]{\label{Fig:mult_dep}\includegraphics[scale=0.32]{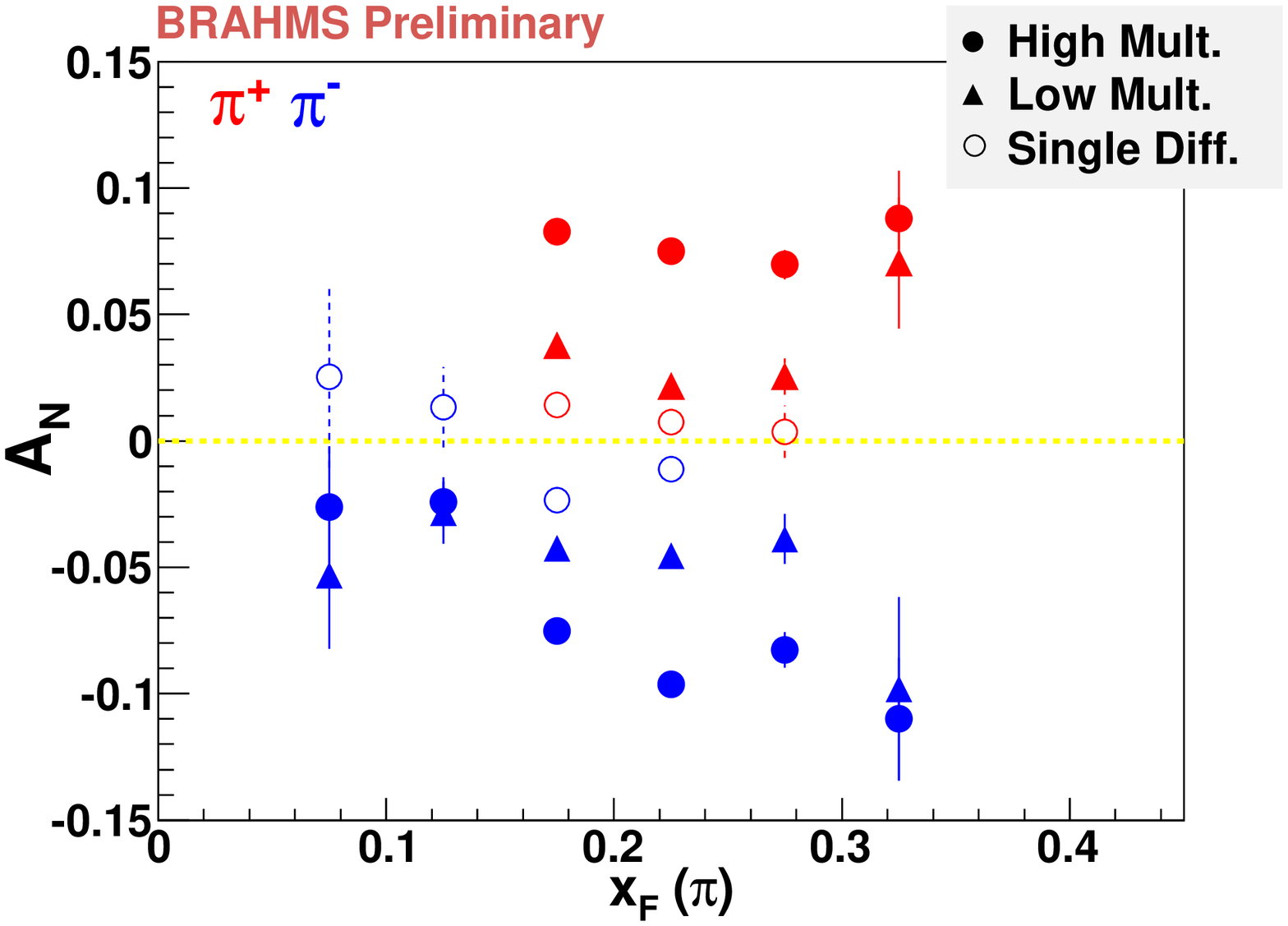}} \\
  \end{center}
\vskip -0.9cm
  \caption{(a) Ratio of the $p_T$ spectra for charged hadrons at y$\sim$0.  Spectra for the 
lowest-multiplicity (solid circles) 
and middle-multiplicity (circles) are divided by the spectrum for the highest-multiplicity. 
(b) Multiplicity-dependent $A_N$($\pi^{\pm}$) for the highest-multiplicity 
(solid circle), lowest-multiplicity (solid triangles), and single diffractive events (circles).} 
  \label{Fig:mult}
\end{figure}
\section{Multiplicity-dependent Single Spin Asymmetries}
Particle multiplicity in $p+p$ carries information 
on the collision dynamics, and  is expected to be sensitive to 
the impact parameter and ``hardness'' of the collisions. 
Since $\sim$10\%~\cite{star_sys} 
of $dN/d\eta$ near mid-rapidity estimated to be originated from hard processes in $p+p$ at $\sqrt{s}=200 GeV$, 
spectra for high multiplicity events are expected to be hard, i.e. have higher $\langle p_T \rangle$.
Multiplicity of the event was selected utilizing the Tile Multiplicity Array (TMA)~\cite{tma}, which measured
charged particles in $-2 < \eta < 2$.   
The data  were divided into three multiplicity classes: lowest, middle, and highest multiplicity events.  
Fig.~\ref{Fig:rpp} shows the ratios of normalized yields of lowest- and middle-multiplicity class events divided by
the highest-multiplicity events  reconstructed by Mid-Rapidity Spectrometer (MRS)~\cite{brahms_nim} at $y$$\sim$0. 
The ratio of lowest/highest multiplicity as function of $p_T$ is very similar as what has been observed by 
STAR~\cite{lisa} of 1-track/9-track events. 

Figure~\ref{Fig:mult_dep} shows SSAs for the highest-multiplicity and the lowest-multiplicity events. 
The figure also shows SSAs for the highly single diffractive enriched process.   
Single diffractive events are from a process  with a large rapidity gap between the backward beam rapidity and the forward
rapidities characterized by excluding events with a hit(s) in backward ``CC'' detector~\cite{prl} which measured 
charged particles in $-$3.25 $>$ $\eta$ $>$ $-$5.25 and also no hits in TMA.
The multiplicity dependent SSAs show that events with more produced particles give stronger asymmetries for pions.
The strong dependence of pion SSAs on the multiplicity can imply that the asymmetries are likely originated from hard collisions, 
but also suggests that there might be significant non-pQCD effects in play such as the collision geometry and 
the energy conservation in the collision process.
\section{Collision Energy-dependent Single Spin Asymmetries}
\begin{figure}[htp]
  \begin{center}
    \subfigure[]{\label{Fig:e704}\includegraphics[scale=0.50]{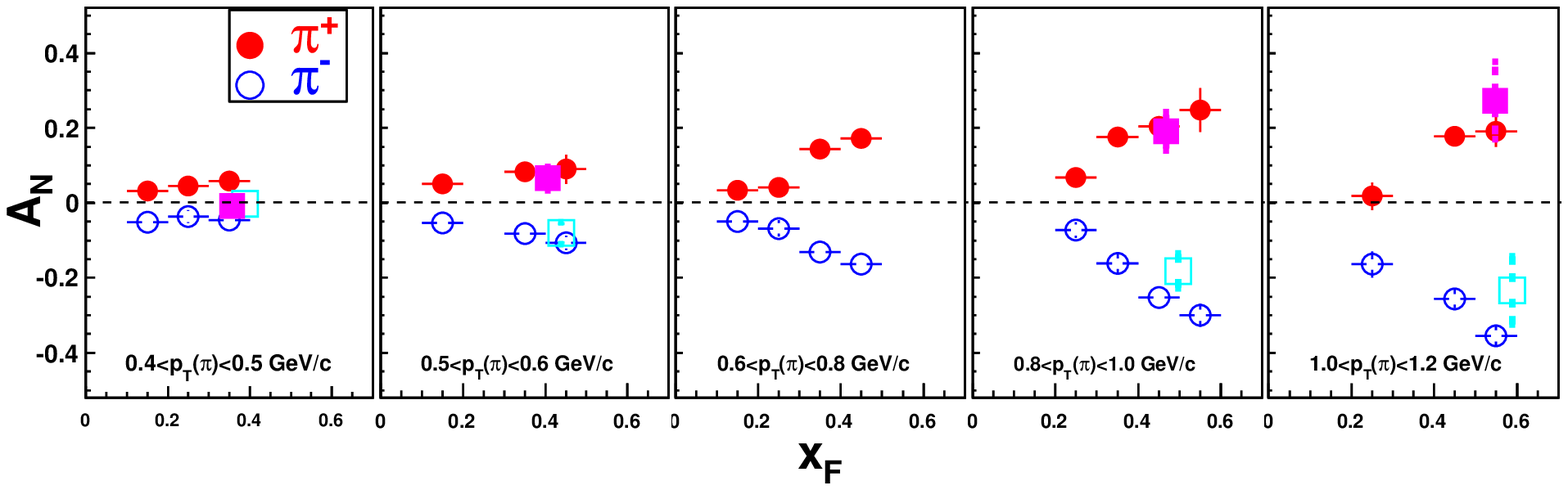}}
    \subfigure[]{\label{Fig:62_200}\includegraphics[scale=0.19]{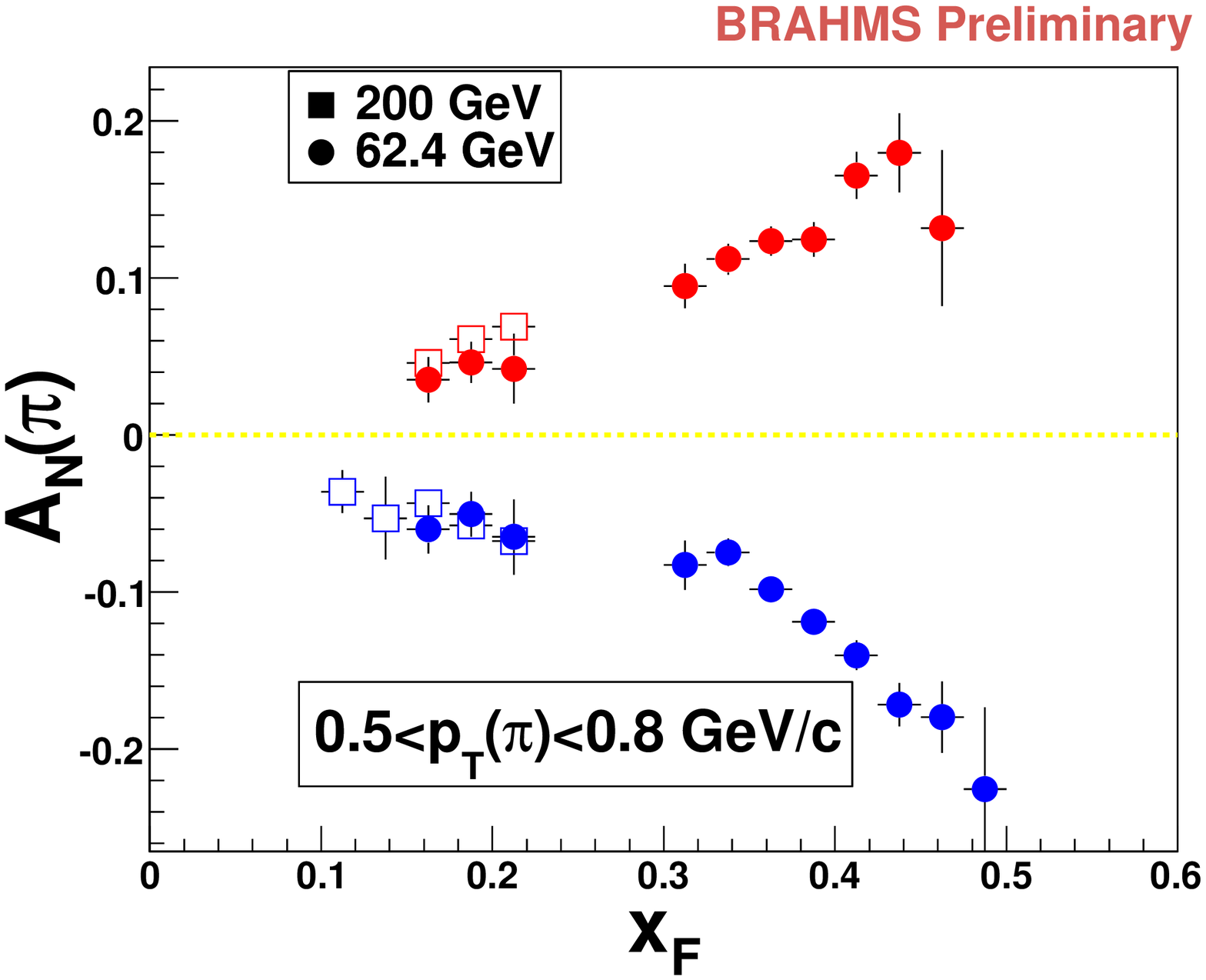}} \\
  \end{center}
\vskip -0.9cm
  \caption{(a) $A_N$  vs. $x_F$  for $\pi^+$ and $\pi^-$ at $\sqrt{s}$ = 62.4 GeV (circles)~\cite{prl} 
at fixed $p_T$ values of 
$0.4<p_T<0.5$, $0.5<p_T<0.6$, $0.6<p_T<0.8$, $0.8<p_T<1.0$,  and $1.0<p_T<1.2$ GeV/$c$, and compared
with $A_N$ at $\sqrt{s}$ = 19.4 GeV (squares)~\cite{bravar}. (b) Comparisons of $A_N(\pi^{\pm})$ at $\sqrt{s}$=62.4 and 200 GeV
at 0.5$<$$p_T$$<$0.8 GeV/$c$.} 
\label{Fig:energy}
\end{figure}

As the collision energy increases, more pQCD applicable processes
are expected to be dominating the mechanism responsible for generating transverse SSAs. 
Energy dependent asymmetry measurements are then expected to provide 
some insight on how the mechanism responsible for SSAs changes.  
Figure~\ref{Fig:e704} shows comparison of charged pion asymmetries measured at $\sqrt{s} =$ 19.4~\cite{adams} 
and 62.4 GeV~\cite{prl} and
Fig.~\ref{Fig:62_200} shows two measurements at  62.4 GeV~\cite{prl} and 200 GeV where the two measurements 
kinematically overlaps 
at 0.5 $<$ $p_T(\pi)$ $<$ 0.8 GeV/$c$. The asymmetries and their
$x_F$-dependence are qualitatively in agreement with the measurements from E704/FNAL~\cite{adams}.
It is noted that the $p_T$ range for the comparison
might be too low for applying pQCD.
Contrary to the observation that there is no significant energy dependence in SSAs for pions in $p(\bar p)+p$ as shown in 
Fig.~\ref{Fig:energy}, significantly different SSAs have been measured depending on the energy in SIDIS by 
HERMES~\cite{hermes} and COMPASS~\cite{compass}.
The energy independence of SSAs in $p(\bar p)+p$ indicates that transverse SSAs are unlikely dominated by 
partonic processes at the measured kinematic region 
and not mainly driven by a process independent factorized process. 

\section{Summary}
Transverse SSAs and cross-sections 
for inclusive identified charged hadron production at forward rapidities in $p$$^\uparrow$+$p$
at $\sqrt{s}$ = 200 GeV have been measured in BRAHMS at RHIC.
The cross-section measurements show that the energy regime is pQCD applicable 
as the spin-averaged cross-section data were 
described by the pQCD inspired models in a wide range of rapidity. The measured
$p_T$, flavor, multiplicity and energy dependent SSAs of identified hadrons 
suggest the manifestation of non-pQCD phenomena in the mechanism driving SSAs.
The differential SSA results allow more complete and stringent tests of 
theoretical models describing spin degree of freedom 
in the context of pQCD and non-pQCD in the RHIC energy regime.


\begin{footnotesize}


%

\end{footnotesize}


\end{document}